\documentclass[aps,
reprint,onecolumn,
preprintnumbers,
nofootinbib,
superscriptaddress,
amsmath,amssymb,aps]{revtex4-2}
\usepackage[utf8]{inputenc}
\usepackage{mathrsfs}

\usepackage{array}
\usepackage{ragged2e}

\usepackage{graphicx}
\graphicspath{{./figs/}}
\usepackage{dcolumn}
\usepackage{bm}

\usepackage{mathrsfs}

\usepackage[normalem]{ulem}

\usepackage{enumitem}

\usepackage{bm}

\usepackage{color}
\usepackage[dvipsnames, table]{xcolor}
\definecolor{blue}{rgb}{0,0,0.5}
\definecolor{lightgray}{gray}{0.95} 

\usepackage[colorlinks,linkcolor=blue,urlcolor=blue,citecolor=blue]{hyperref}
\usepackage{graphicx}
\usepackage{slashed}
\usepackage{enumitem}
\usepackage{bm}
\usepackage{physics}
\usepackage{xspace}
\usepackage[T1]{fontenc}
\usepackage{lmodern}
   
\usepackage{tabularx} 
\usepackage{booktabs}  
\usepackage{diagbox}
\usepackage{multirow}

\newcommand{\be}{\begin{equation}}
\newcommand{\ee}{\end{equation}}
\newcommand{\bea}{\begin{eqnarray}}
\newcommand{\eea}{\end{eqnarray}}

\newcommand{\mc}{\mathcal}

\newcommand{\noi}{\noindent}
\def\mcB{\mc B}

\newcommand{\GeV}{{\rm GeV}}

\newcommand{\BDTt}{{\rm BDT}_2}

\def\reffig#1{Fig.~\ref{#1}}
\def\refsec#1{Sec.~\ref{#1}}

\def\refcite#1{Ref.~\cite{#1}}
\newcommand{\Ba}{\mc B_a}
\newcommand{\qqrec}{q^2_{\rm rec}}
\def\BKvv{B^+ \! \to \! K^+\nu\bar\nu}
\def\BKa{B^+ \! \to \! K^+ a}

\def\nunu{\nu \bar \nu}
\newcommand{\muvv}{\mu_{\nunu}}

\newcommand{\FVsb}{(F_V)_{sb}}

\newcommand{\us}{Abumusabh:2025zsr}
\newcommand{\gaertner}{Gartner:2026clx}
\newcommand{\abuagnostic}{Belle-II:2025lfq}
\newcommand{\camalich}{MartinCamalich:2020dfe}

\newcommand{\pdg}{ParticleDataGroup:2024cfk}

\newcommand{\BelleIIevi}{Belle-II:2023esi}
\newcommand{\fridell}{Fridell:2023ssf}
\newcommand{\boltonI}{Bolton:2024egx}
\newcommand{\boltonII}{Bolton:2025fsq}
\newcommand{\ferber}{Ferber:2022ewf}

\newcommand{\beal}{\begin{aligned}}
\newcommand{\eeal}{\end{aligned}}

\makeatletter
\g@addto@macro\bfseries{\boldmath}
\makeatother

\makeatletter
\def\p@subsection{}
\makeatother

\newlist{todolist}{itemize}{2}
\setlist[todolist]{label=$\square$}
\usepackage{pifont}
\DeclareOldFontCommand{\rm}{\normalfont\rmfamily}{\mathrm}
\DeclareOldFontCommand{\sf}{\normalfont\sffamily}{\mathsf}
\DeclareOldFontCommand{\tt}{\normalfont\ttfamily}{\mathtt}
\DeclareOldFontCommand{\bf}{\normalfont\bfseries}{\mathbf}
\DeclareOldFontCommand{\it}{\normalfont\itshape}{\mathit}
\DeclareOldFontCommand{\sl}{\normalfont\slshape}{\@nomath\sl}
\DeclareOldFontCommand{\sc}{\normalfont\scshape}{\@nomath\sc}

\newcommand{\bof}[1]{\smallskip \noindent {\bfseries #1}~--~ \noi}
\newcommand{\noibf}[1]{\noindent {\bfseries #1}~}

\begin{document}

\preprint{LAPTH-033/26}

\title{The $B^+ \to K^+ \nu \bar \nu$ decay as a QCD axion search:\\[0.2cm]
comparing reinterpretation approaches}

\author{Merna Abumusabh}

\email{merna.abumusabh@iphc.cnrs.fr}

\affiliation{%
{\itshape IPHC, Universit\'{e} de Strasbourg et CNRS, 67200 Strasbourg, France}
}%

\author{Giulio Dujany}

\email{giulio.dujany@iphc.cnrs.fr}

\affiliation{%
{\itshape IPHC, Universit\'{e} de Strasbourg et CNRS, 67200 Strasbourg, France}
}%

\author{Diego Guadagnoli}

\email{diego.guadagnoli@lapth.cnrs.fr}

\affiliation{%
{\itshape LAPTh, Universit\'{e} Savoie Mont-Blanc et CNRS, 74941 Annecy, France}
}%

\author{Méril Reboud}

\email{meril.reboud@ijclab.in2p3.fr}

\affiliation{%
{\itshape IJCLab, Pôle Théorie, et Université Paris-Saclay, 91400 Orsay, France}
}%

\author{Claudio Toni}

\email{claudio.toni@lapth.cnrs.fr}

\affiliation{%
{\itshape LAPTh, Universit\'{e} Savoie Mont-Blanc et CNRS, 74941 Annecy, France}
}%

\begin{abstract}

\noi Two recent independent analyses of Belle~II $\BKvv$ data yield limits on $\mcB(\BKa)$---the two-body mode to a light invisible particle such as the QCD axion---differing by a factor of roughly four; we trace this to the choice of kinematic variable space. The central figure of merit is the resolution in the reconstructed di-neutrino invariant mass $\qqrec$: fine-grained binning resolves the narrow axion signal, while coarse binning dilutes it into a background-dominated range. A BDT axis trained on $\BKvv$ adds little discriminating power for $\BKa$, as this axis is largely uncorrelated with $q^2$.

These expectations are confirmed by a set of numerical tests. The subleading shape systematics omitted from our $\qqrec$-based approach {\em lower}, not raise, the $\BKa$ limit: by better accommodating the $\BKvv$ shape, they leave less room for the axion signal, making our $\qqrec$-based bound conservative, if anything. 
A dedicated reanalysis confirms that the kinematic-axes choice alone accounts for the factor-of-four sensitivity difference, and that the $\BKa$ bound varies sizeably within the $\qqrec\times\eta(\BDTt)$ space, depending on the SM-likeness of $\BKvv$, thus losing the dual-probe feature of our $\qqrec$-based approach.

These results point to a broader consideration: likelihoods dominated by BDT variables are of limited use for reinterpretations when the signal shape differs appreciably from the BDT's training signal. We therefore advocate that experimental collaborations publish likelihood projections in physical variable spaces alongside BDT-based likelihoods, to maximise the reinterpretability of their measurements.

\end{abstract}

\maketitle

\section{Introduction}
\label{sec:intro}

\noi The reinterpretation of Belle~II $\BKvv$ data as a search for the two-body decay $\BKa$, where $a$ is a light invisible state such as the QCD axion or a generic axion-like particle (ALP), has recently been addressed by two independent groups.
The first approach~\cite{\us} introduces a model-independent framework using only public data, analytically reconstructing the mapping between true and reconstructed kinematic variables within the statistically dominant Inclusive Tagging Analysis (ITA). Applied to $\BKa$, this yields the strongest existing bounds on $\mcB(\BKa)$ and on the coupling-rescaled Peccei--Quinn scale $|\FVsb|$, improving previous limits by about one order of magnitude~\cite{\camalich,\ferber}.
A second approach~\cite{\gaertner} is based on the published model-agnostic likelihood of Ref.~\cite{\abuagnostic}. For the physically motivated QCD axion, the resulting limits on $\mcB(\BKa)$ are weaker by a factor of roughly four than those of the first approach.

In this note we discuss the origins of this difference. Both approaches begin from {\em projections} of the same underlying Belle~II data and simulation: neither has access to the full unpublished event-level information. The relevant difference lies in the granularity and the kinematic axes of these projections. The likelihood of \refcite{\abuagnostic} is a function of two variables: a coarsely binned $\qqrec$ (three bins over the full kinematically accessible range), and the transformed $\BDTt$ output $\eta(\BDTt)$, a variable designed to classify events as signal-like for $\BKvv$. As we argue below, neither variable is well matched to a narrow invisible signal: the coarse $\qqrec$ binning washes out the structured signal distribution, while $\eta(\BDTt)$ carries little or no discriminating power between $\BKa$ and background.
By contrast, \refcite{\us} works directly with the fine-grained $\qqrec$ distribution of Fig.~17 of \refcite{\BelleIIevi}, which retains approximately seven times more $\qqrec$ resolution and is the natural variable for narrow-signal recasting.

The rest of this note is organized as follows. In \refsec{sec:origins} we develop the analytical basis for the expected sensitivity difference, discussing in detail the role of $\qqrec$ granularity, the properties of the $\eta(\BDTt)$ axis, and a formal limitation of the reweighting method of \refcite{\abuagnostic} for delta-function signals. In \refsec{sec:robustness} we subject the approach of \refcite{\us} to a series of robustness tests, varying the background-normalization uncertainty and examining the profile likelihood for $\BKvv$; these tests simultaneously validate our uncertainty estimate and clarify the physics of the shape separation between the $\BKa$ and $\BKvv$ components. In \refsec{sec:gaertnerlike} we construct an independent analysis in the $\qqrec\times\eta(\BDTt)$ space, directly comparable to \refcite{\gaertner}, and use it to quantify the sensitivity loss attributable to the choice of kinematic axes. \refsec{sec:conclusions} collects our conclusions.

\section{Origins of the sensitivity difference} \label{sec:origins}

The starting point of \refcite{\us} is Fig.~17 of \refcite{\BelleIIevi}, which displays the $\qqrec$ distribution in the ITA signal region with approximately 21 bins over the range $[-1,25]~\GeV^2$, derived directly from Belle~II data and simulation. The starting point of \refcite{\gaertner} is Fig.~1 of \refcite{\abuagnostic}, which encodes the same observable in the binning of the published likelihood: $4\times3$ bins in $\eta(\BDTt)\times\qqrec$, with the three $\qqrec$ bin boundaries at $\{-1,4,8,25\}~\GeV^2$.
Both starting points are projections of the same underlying Belle~II data and simulation, which exist in a richer, higher-dimensional form than either figure captures. Fig.~17 of \refcite{\BelleIIevi} is a projection of actual data and internal simulation; Fig.~1 of \refcite{\abuagnostic} is a projection of the same underlying data encoded in the published likelihood. Neither is the ``ultimate truth''; the difference between them is one of $\qqrec$ resolution, approximately seven times finer in our case.

\bof{Resolution in $\qqrec$: the key figure of merit for narrow signals}%
This difference in resolution is consequential specifically for narrow invisible signals. For the QCD axion ($m_a\simeq 0$), the signal is monochromatic at $q^2=0$ and maps, via the kinematic relation derived analytically in \refcite{\us}, onto a distribution in $\qqrec$ that is broadened by the unmeasured decay angle but remains structured and concentrated near $\qqrec\simeq 0$. With the published likelihood's three $\qqrec$ bins, this signal is subsumed into the single bin $[-1,4]~\GeV^2$, where it competes with the $\BKvv$ background that populates the same bin. With approximately 21 bins, the signal structure is resolved over several bins, each carrying a smaller background contribution, and sensitivity improves proportionally. For the $\BKvv$ measurement itself---for which the three-bin structure was designed---the two approaches are equivalent: the $\nunu$ spectrum is smooth and broad, and its integral over any bin is well represented at either resolution. The asymmetry between the two cases is therefore not a limitation of the $\BKvv$ analysis, but a structural feature of repurposing it for narrow signals.\footnote{The principle of using fine-grained $\qqrec$ binning for narrow-signal recasting has also been consistently exploited in Refs.~\cite{\fridell,\boltonI,\boltonII}, and has shown value in the reinterpretation of BaBar's differential $s_B$ distribution, performed bin-by-bin in \refcite{\ferber}.}

\bof{The $\eta(\BDTt)$ axis: a variable optimized for $\BKvv$}%
The second axis of the published likelihood, $\eta(\BDTt)$, raises an independent concern of the same nature. The variable $\BDTt$ was trained to separate $\BKvv$ signal from background: a value $\eta\simeq 1$ identifies events as signal-like for $\nunu$, not for $\BKa$. The two signals differ kinematically: the axion produces a monochromatic kaon in the $B$-meson rest frame, whereas $\nunu$ produces a continuous kaon spectrum. As a result, the $\eta(\BDTt)$ axis may carry little or no discriminating power between $\BKa$ and background, and may even preferentially suppress axion signal events that are not $\nunu$-like. More precisely, the sole variable that truly discriminates $\BKa$ from $\BKvv$ is $q^2$ itself---and $\eta(\BDTt)$ is only loosely correlated with $q^2$, since the BDT was not trained to exploit this distinction. The 2D space of \refcite{\abuagnostic} is therefore not a neutral basis for a model-independent recast of narrow signals: it is a space optimized for a specific, and different, signal topology.

\bof{The reweighting method in the delta-function limit}%
The two limitations described above---coarse $\qqrec$ binning and a BDT axis tuned to $\nunu$ kinematics---are both natural consequences of applying a likelihood designed for $\BKvv$ to the qualitatively different case of a narrow two-body signal. They are also directly connected to a formal property of the reweighting method of \refcite{\abuagnostic}. That method constructs alternative signal templates via $w(q^2)=\sigma_1(q^2)/\sigma_0(q^2)$, where $\sigma_0$ is the smooth SM $\nunu$ spectrum, and notes explicitly that it ``performs best when the alternative theory remains close to the null distribution'', warning that large deviations produce ``very large weights'' and ``potentially unreliable results''. For a narrow or zero-width invisible state, $\sigma_1(q^2)\propto\delta(q^2-m_a^2)$, which is by construction maximally far from the smooth null distribution. This is precisely the least favorable regime for the method. The resulting large weights, concentrated in a single $q^2$ bin, compound the sensitivity loss from the coarse $\qqrec$ binning: the signal information is already diluted at the level of the kinematic axes, and the reweighting step cannot recover it.

\bof{The role of systematic uncertainties}%
A separate question is whether the factor-of-four difference in the $\mcB_a$ limits between the two approaches might be driven, at least in part, by the more conservative treatment of uncertainties in \refcite{\gaertner}, which incorporates the full set of systematic uncertainties from the Belle~II analysis, whereas \refcite{\us} retains only the dominant background-normalization component. We argue that this is not the case. The background-normalization uncertainty accounts for approximately 50\% of the total systematic uncertainty in the ITA~\cite{\BelleIIevi}, and the remaining components individually contribute at a level well below what would be needed to explain a factor-of-four degradation. Moreover, the inclusion of additional systematic components tends to {\em lower} the $\mcB_a$ limit rather than raise it, because it gives the fit more freedom to accommodate the $\BKvv$ shape and thereby leaves less room for the $\BKa$ signal. Therefore, the neglect of subleading shape systematics makes the $\mcB_a$ bound from \refcite{\us}, if anything, slightly conservative. These conclusions are verified numerically in \refsec{sec:robustness}.

\section{Robustness of the approach in \refcite{\us}} \label{sec:robustness}

\noi The analysis of \refcite{\us} incorporates one dominant systematic uncertainty: the normalization of the total background, estimated at approximately 1\% for the ITA.  This estimate is derived from the pull distribution of Fig.~17 of \refcite{\BelleIIevi}. Identifying the pull in each $\qqrec$ bin $i$ as $|n^i_{\rm obs} - n^i_{\rm hst}| / \sqrt{n^i_{\rm obs} + (\sigma^i_{\rm hst})^2}$, where $n^i_{\rm obs}$ and $n^i_{\rm hst}$ are the observed yield and the stacked-histogram prediction respectively, and inverting this relation for $\sigma^i_{\rm hst}$, one finds that the relative uncertainty $\sigma^i_{\rm hst}/n^i_{\rm hst}$ is approximately constant across bins at the level of $\sim 1\%$. This motivates modeling the background uncertainty as a single overall normalization factor, with a relative uncertainty of 1\%, common to all bins.

In this section we subject this choice to a series of tests by varying the background-normalization error over a wide range and examining the effect on two separate quantities: the $\mcB_a$ limit and the profile likelihood for $\BKvv$. The two quantities respond differently to this variation, and the contrast between them is itself informative. The $\mcB_a$ signal is concentrated near $\qqrec\simeq 0$ and has a qualitatively different shape from both the $\BKvv$ signal and its backgrounds; the data in all bins but one therefore constrain the $\BKvv$ component independently of the normalization prior, leaving the $\mcB_a$ limit unaffected. The $\BKvv$ profile likelihood, by contrast, is sensitive to the normalization error, because the $\BKvv$ signal and its backgrounds have similar shapes in $\qqrec$ and are therefore not shape-separable. These expectations are confirmed by the tests described below, which simultaneously validate the 1\% error estimate and clarify the physical basis for the dual-probe property specific to the approach in~\refcite{\us}.

Throughout this paper, limits are quoted at 95\% CL and restricted to the ITA-only analysis, to allow direct numerical comparison with \refcite{\gaertner}.
The definitive 90\% CL bounds, following PDG convention~\cite{\pdg}, are those of \refcite{\us} and are to be found therein.

\begin{figure}[t]
\begin{center}
\includegraphics[scale=0.65]{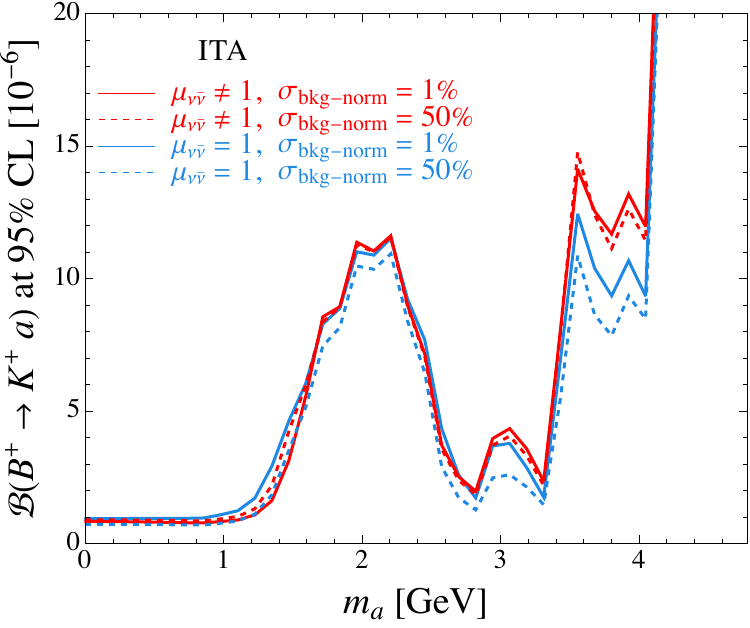} \hfill
\includegraphics[scale=0.75]{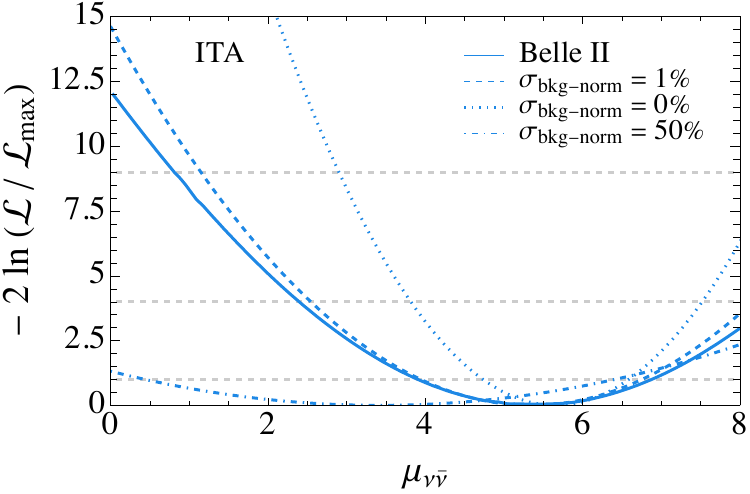}
\caption{%
Effect of varying the background-normalization error on two observables. 
{\em Left.} $\Ba$ limit as a function of $m_a$: solid (dashed) lines use the reference 1\%~\cite{\us} (inflated 50\%) error; blue and red distinguish $\muvv=1$ and $\muvv\neq 1$. 
{\em Right.} Likelihood profile as a function of $\muvv$ with $\Ba=0$, for errors of 0\%, 1\%, and 50\%, compared to the Belle~II result of \refcite{\BelleIIevi} (see legend and text for details).
}
\label{fig:bkg_err_dependence}
\end{center}
\end{figure}

\medskip
{\bf Test 1: Insensitivity of $\mcB_a$ bound from background-normalization error size~--~}%
\noi We first verify the robustness of the $\mcB_a$ limit by repeating the maximization of the likelihood of \refcite{\us} with the {\em absolute} background-normalization error inflated from 1\% to as much as 50\%.
The resulting $\mcB_a$ limits as a function of $m_a$ are shown in \reffig{fig:bkg_err_dependence} (left panel), where solid lines correspond to the 1\% case and dashed lines to the 50\% case, with red and blue denoting $\muvv\neq 1$ and $\muvv=1$ respectively.
Here $\muvv\equiv\mcB(\BKvv)/\mcB_{\rm SM}(\BKvv)$ parametrizes possible new-physics contributions to the $\BKvv$ amplitude. The four curves are in excellent mutual agreement throughout the full $m_a$ range.

Two conclusions follow. First, the $\mcB_a$ limit is insensitive (cf. solid vs. dashed) to the assumed background-normalization error, confirming that the data themselves constrain the $\BKvv$ component in all bins but one, irrespective of the prior. Second, this insensitivity holds for {\em both} choices of $\mu_{\nunu}$, and the two choices (cf. red vs. blue) themselves yield near-identical limits---a direct numerical manifestation of the dual-probe property established analytically in \refcite{\us}: the $\mcB_a$ and $\mcB_{\nunu}$ bounds are, within the approach of \refcite{\us}, statistically decoupled to an excellent approximation, and this decoupling is robust against the dominant systematic.

\medskip
{\bf Test 2: Dependence of $\mcB_{\nunu}$ profile likelihood on background-normalization error size~--~}%
We next set $\mcB_a = 0$ and examine the profile likelihood as a function of $\mu_{\nunu}$, repeating the exercise of Fig.~4 of \refcite{\us} with background-normalization errors of absolute sizes of 0\%, 1\%, and 50\%. The results are shown in \reffig{fig:bkg_err_dependence} (right), alongside the Belle~II experimental profile of Fig.~16 of \refcite{\BelleIIevi} for comparison. The three cases bracket the experimental profile in a diagnostically clean way: the 0\% error yields a profile that is too sharp, the 50\% error yields one that is too broad and loses the $3\sigma$ evidence for $\BKvv$, and the 1\% error reproduces the experimental profile closely. This sensitivity to the normalization error contrasts sharply with the insensitivity observed for $\mcB_a$ in Test 1, and its origin is transparent: {\em the $\BKvv$ signal and its backgrounds have similar shapes in $\qqrec$}, so that only their overall normalization distinguishes them. Any over- or underestimate of the normalization uncertainty therefore propagates directly into the curvature of the $\mu_{\nunu}$ profile. The fact that 1\% reproduces the experimental profile was already noted in \refcite{\us} and constitutes an independent, a posteriori validation of the error estimate itself---derived in \refcite{\us} from the pull distribution of Fig.~17 of \refcite{\BelleIIevi}.

\begin{figure}[t]
\begin{center}
\includegraphics[scale=0.75]{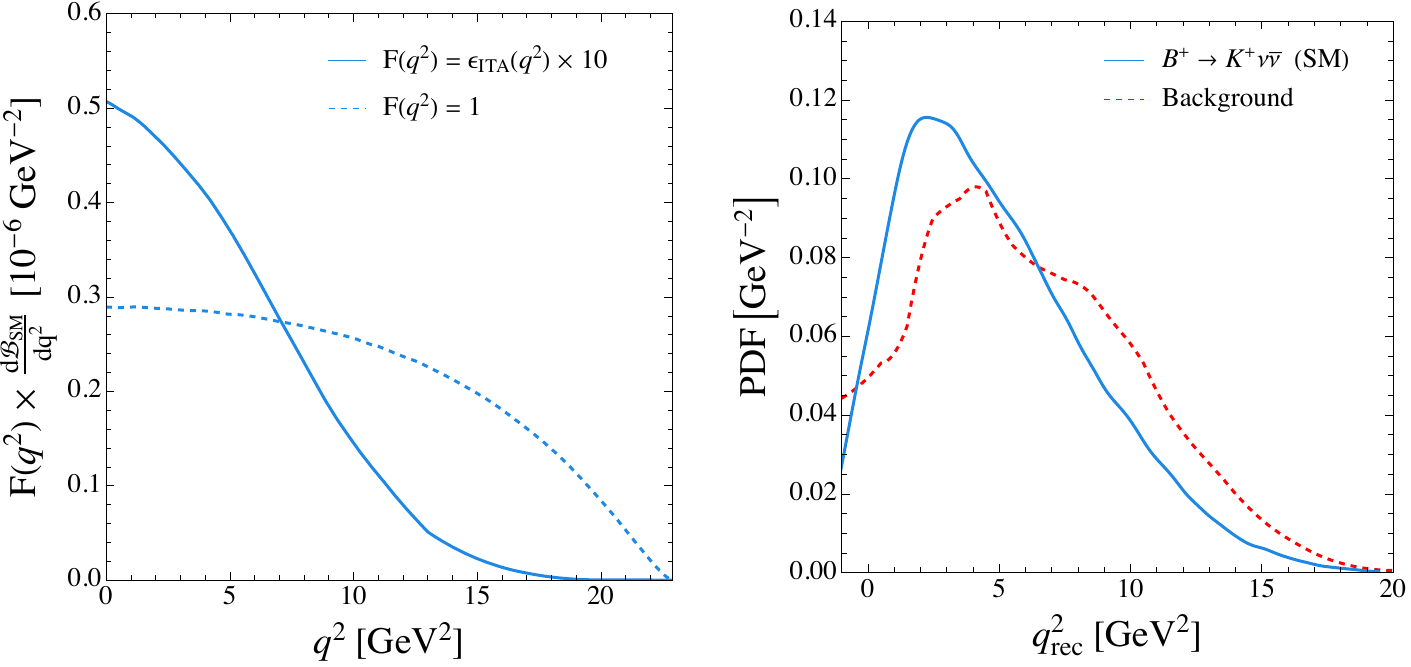} \hfill
\caption{%
Monte Carlo illustration of the $\BKvv$ signal shape.
{\em Left.} Physical differential branching fraction $d\mcB_{\rm SM}/dq^2$ (dashed) and its product with the ITA efficiency $\varepsilon_{\rm ITA}(q^2)$ (solid, magnified by a factor of 10), as functions of $q^2$: both are non-zero and maximal at $q^2=0$ and decrease monotonically toward the kinematic endpoint.
{\em Right.} The $\BKvv$ signal yield as a function of $\qqrec$ (solid, blue), obtained from the product in the left panel via the $q^2\to\qqrec$ kinematic mapping of \refcite{\us}, compared to the stacked background distribution of \refcite{\BelleIIevi}. Both curves are normalized to unity; the kinematic smearing redistributes events away from $\qqrec\simeq 0$, producing a peak around $\qqrec\simeq 4$--$5~\GeV^2$ for both distributions.
}
\label{fig:BKvv_shape}
\end{center}
\end{figure}
The shape similarity invoked in the previous paragraph can be verified directly. The stacked background distribution visible in Figs.~17--18 of \refcite{\BelleIIevi} is small at low $\qqrec$, peaks around $\qqrec\simeq 4$--$5~\GeV^2$, and falls off above.
The $\BKvv$ signal yield as a function of $\qqrec$ is shaped by three ingredients: the SM differential branching fraction $d\mcB/dq^2$, which is non-zero and maximal at $q^2=0$ and decreases monotonically, as shown by the dashed line in the left panel of 
\reffig{fig:BKvv_shape}~(left); the ITA selection efficiency $\varepsilon_{\rm ITA}(q^2)$, which is likewise maximal near $q^2\simeq 0$ and falls steeply; and the $q^2\to\qqrec$ kinematic mapping, which smears each $q^2$ value over a range of $\qqrec$ values via the unmeasured polar angle $\theta^{**}$ of the signal kaon in the $B$-meson rest frame.
The product $\varepsilon_{\rm ITA}\times d\mcB/dq^2$, shown as a solid line in \reffig{fig:BKvv_shape}~(left), magnified by a factor of 10, inherits the same monotonically decreasing shape. 
The right panel of \reffig{fig:BKvv_shape} shows the corresponding yield as a function of $\qqrec$: the kinematic smearing redistributes events away from $\qqrec\simeq 0$, producing the zero-then-peak-then-decrease shape at $\qqrec\simeq 4$--$5~\GeV^2$ that closely mirrors the stacked background distribution of \refcite{\BelleIIevi}, which is also shown in the right panel of \reffig{fig:BKvv_shape} for comparison.

This explains why the $\BKvv$ profile likelihood depends on the background-normalization error: in the absence of shape discrimination, the normalization prior is the primary handle on the separation between signal and background. It also explains, by contrast, why the $\mcB_a$ limit does {\em not} depend on it: the axion signal, concentrated near $\qqrec\simeq 0$, has a qualitatively different shape from both the $\BKvv$ signal and its backgrounds, and is therefore shape-separated from them regardless of the normalization prior.

\section{\boldmath A reanalysis in the $q^2_{\rm rec}\times\eta(\BDTt)$ space} \label{sec:gaertnerlike}
\def\usGaertnerlike{q^2_{\rm rec}\times{\rm BDT}}

\noi This section probes the sensitivity difference between the two approaches through two complementary analyses. The first is a direct rerun of the published \texttt{pyhf} likelihood of \refcite{\gaertner}, which establishes a numerical baseline for the \refcite{\gaertner} result and quantifies the roles of $\muvv$ freedom and of the shape systematics within that framework. The second is a purpose-built analysis---referred to hereafter as the $\usGaertnerlike$ analysis---that applies the likelihood framework of \refcite{\us} to the same starting point used by \refcite{\gaertner}: the $\qqrec\times\eta(\BDTt)$ histogram of Fig.~15 of \refcite{\BelleIIevi}, rather than the fine-grained $\qqrec$ distribution of Fig.~17. The statistical method, signal model, and treatment of the background normalization are otherwise identical to those of \refcite{\us}. Any difference in the $\mcB_a$ limit between \refcite{\us} and the $\usGaertnerlike$ analysis is therefore directly attributable to the choice of kinematic axes, while any residual difference between the $\usGaertnerlike$ analysis and \refcite{\gaertner} reflects the additional systematic components present in the latter but absent in both the former and \refcite{\us}.

\medskip
{\bf Test 3: Direct rerun of the \refcite{\gaertner} likelihood~--~}%
We run the published \texttt{pyhf} likelihood of \refcite{\gaertner} directly, as released on \texttt{HEPData}~\cite{\abuagnostic}, incorporating the full set of 231 nuisance parameters of the Belle~II analysis. We implement the same narrow-ALP signal model as \refcite{\gaertner} and consider three variants:
\def\MuGauAllShapes{{\em (i)}}
\def\MuGauNormOnly{{\em (ii)}}
\def\MuFreeAllShapes{{\em (iii)}}
\def\MuGauAllShapesNUM{4.0 \times 10^{-6}}
\def\MuGauNormOnlyNUM{4.8 \times 10^{-6}}
\def\MuFreeAllShapesNUM{1.81 \times 10^{-6}}
\begin{enumerate}[label={\em (\roman*)},noitemsep]
\item $\muvv$ constrained as a Gaussian around 1 with 5\% width, all 231 shape nuisances free---reproducing \refcite{\gaertner} exactly;
\item same $\muvv$ constraint, all shape nuisances fixed to nominal values except the 
background-normalization component---isolating the role of shape systematics;
\item $\muvv$ entirely free, all shape nuisances free---probing the effect of the $\muvv$ constraint itself.
\end{enumerate}

\begin{figure}[t]
\begin{center}
\includegraphics[scale=0.58]{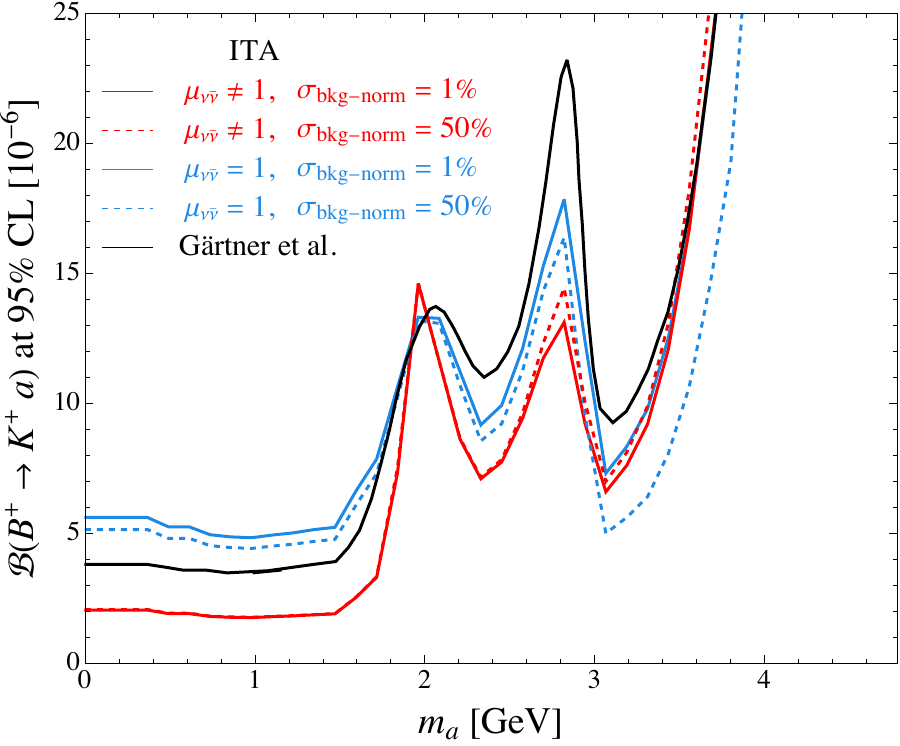}
\includegraphics[scale=0.58]{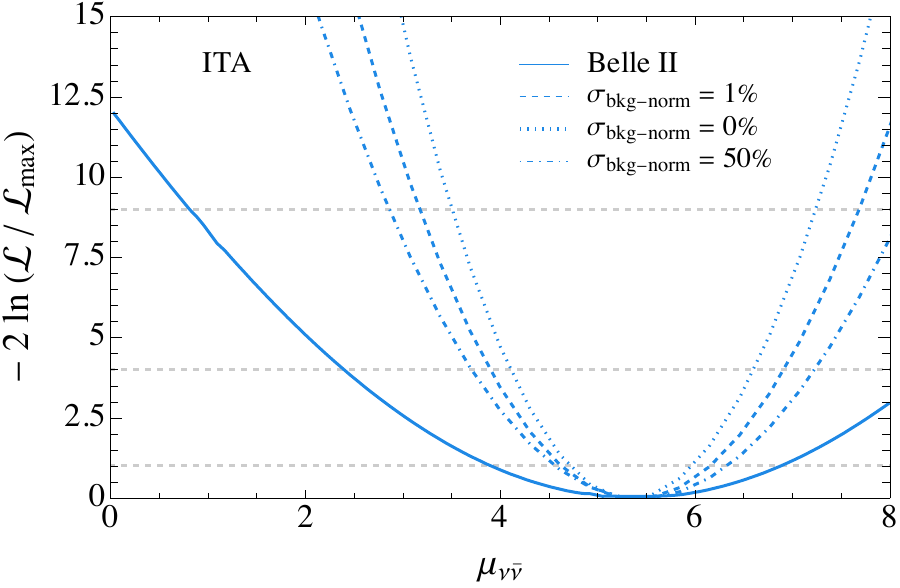} \hfill
\caption{%
{\em Left.} $\Ba$ limit as a function of $m_a$ obtained from the $\usGaertnerlike$ analysis applied to the ITA dataset. Within this analysis, we consider the cases $\muvv = 1$ (blue lines) and $\muvv \neq 1$ (red lines), as well as a background-normalization error of 1\% (solid) or 50\% (dashed). The black line reproduces, for comparison, the bound obtained in \refcite{\gaertner}.
{\em Right.} Likelihood profile as a function of $\muvv$ with $\Ba=0$ in the $\usGaertnerlike$ analysis, for background-normalization errors of 0\%, 1\%, and 50\% (see legend for line styles). The Belle~II experimental profile of \refcite{\BelleIIevi} is shown for comparison. The three $\usGaertnerlike$ curves are mutually distinguishable (0\% the sharpest, 50\% the broadest) and all remain sharper than the Belle~II profile; see text for discussion.
}
\label{fig:BKa_Gaertnerlike}
\end{center}
\end{figure}

All three results are obtained at $m_a\simeq 0$.
Variant~\MuGauAllShapes\ yields $\MuGauAllShapesNUM$, in precise agreement with the \refcite{\gaertner} result reproduced as the black line in \reffig{fig:BKa_Gaertnerlike}, establishing the numerical anchor point for the $\usGaertnerlike$ analyses of Tests~4 and~5. A hard fix $\muvv=1$ gives an identical result, confirming that the 5\% Gaussian constraint of \refcite{\gaertner} is effectively equivalent to fixing $\muvv=1$.
Variant~\MuGauNormOnly\ yields $\MuGauNormOnlyNUM$: the $\sim20\%$ increase from variant~\MuGauAllShapes\ directly quantifies the role of shape systematics, which lower $\Ba$ by giving the fit more freedom to accommodate the $\BKvv$ shape---consistent with the conclusion of \refsec{sec:origins} that {\em additional systematics act conservatively on $\Ba$}.
Variant~\MuFreeAllShapes\ yields $\MuFreeAllShapesNUM$, close to the red ($\muvv\neq1$) curve of \reffig{fig:BKa_Gaertnerlike} ($\sim2\times10^{-6}$). Together, variants~\MuGauAllShapes\ and~\MuFreeAllShapes\ illustrate a general feature of the $\qqrec\times\eta(\BDTt)$ space: the $\Ba$ limit changes depending on the assumed $\muvv$ by a factor around two---thus losing the ``dual-probe'' property (i.e. the insensitivity of the $\Ba$ bound to the SM-likeness of $\BKvv$) observed in the approach of \refcite{\us}.

The remaining gap between variant~\MuGauNormOnly\ ($\MuGauNormOnlyNUM$) and the blue $\usGaertnerlike$ curve ($\sim5.2$--$5.6\times10^{-6}$) is not attributable to the $\muvv$ treatment, since by the previous discussion both effectively fix $\muvv\approx1$. It instead reflects differences between the two likelihood frameworks: the full \texttt{pyhf} model of \refcite{\gaertner} and our simpler $\usGaertnerlike$ reconstruction from the $\qqrec\times\eta(\BDTt)$ histogram of \refcite{\BelleIIevi}.

\medskip
{\bf Test 4: $\mcB_a$ bound comparison~--~}%
The $\mcB_a$ limits as a function of $m_a$ obtained from the $\usGaertnerlike$ analysis are shown in \reffig{fig:BKa_Gaertnerlike} (left panel) for four combinations of $\muvv$ assumption and background-normalization error: $\muvv$ free or fixed to unity (red vs. blue) and background-normalization error set to the reference value of 1\%, or to 50\% (solid colored vs. dashed). The result from \refcite{\gaertner} (solid black) is also shown for comparison. Three observations follow.

\noibf{Shape reproduction.} First, the $\usGaertnerlike$ analysis neatly reproduces the overall shape of the \refcite{\gaertner} limit as a function of $m_a$---including the plateau at low mass, the local features around $m_a\simeq 2~\GeV$ and $m_a\simeq 2.8~\GeV$, and the rise above $m_a\simeq 3.6~\GeV$---confirming that the $\usGaertnerlike$ analysis captures the essential structure of the \refcite{\gaertner} approach.

\noi {\bf Factor-of-five degradation.} Second, the $\usGaertnerlike$ limit at low $m_a$ ranges from $\sim 2\times10^{-6}$ (red, i.e. $\muvv\neq 1$) to $\sim 5\times10^{-6}$ (blue, i.e. $\muvv=1$), a factor of roughly five above the corresponding limit of \refcite{\us} ($\sim 1\times10^{-6}$), in direct confirmation of the expectation from \refsec{sec:origins} that the coarse $\qqrec$ binning and the signal-agnostic $\eta(\BDTt)$ axis would dilute the structured axion signal distribution and degrade sensitivity.

\noibf{Dependence on background-normalization systematics.} Two further features of the figure are worth noting. The 1\% and 50\% background-normalization error curves are indistinguishable for $\muvv\neq 1$ (red): when $\muvv$ floats freely, the $\BKvv$ component is already optimally accommodated, and the normalization prior plays no additional role. For $\muvv=1$ (blue), a small but visible gap appears between the two: the larger normalization freedom partially compensates for the rigidity of fixed $\muvv$, easing the accommodation of $\BKvv$ and slightly tightening the $\BKa$ limit from $\sim 5.6\times10^{-6}$ (1\%) to $\sim 5.2\times10^{-6}$ (50\%).\footnote{This gap is slightly more pronounced in the $\usGaertnerlike$ analysis than in the analysis of \refcite{\us}, because the coarser $\qqrec$ space provides less intrinsic shape separation.} Both features are fully consistent with the picture developed in \refsec{sec:robustness}.

\noibf{Dependence on residual shape systematics.} Third, the \refcite{\gaertner} limit at low $m_a$ ($\sim 4\times10^{-6}$) sits below the blue $\usGaertnerlike$ curve. We understand this difference as the result of two effects. The dominant one is the inclusion of subleading shape systematics:\footnote{As further confirmation that the gap between the blue $\usGaertnerlike$ curve and \refcite{\gaertner} is not driven by the assumed size of the background-normalization component, inflating the background-normalization error to 200\% leaves the $\usGaertnerlike$ limit essentially unchanged.} as established in Test~3, variant~\MuGauNormOnly\ ($\muvv = {\rm Gaus}(1,5\%)$, background-normalization only) gives $\MuGauNormOnlyNUM$, while variant~\MuGauAllShapes\ ($\muvv = {\rm Gaus}(1,5\%)$, all shape nuisances) recovers the full \refcite{\gaertner} result of $\MuGauAllShapesNUM$, a $\sim20\%$ difference. The secondary factor is the difference between the two likelihood frameworks---\refcite{\gaertner}'s full \texttt{pyhf} model vs. our simpler $\usGaertnerlike$ reconstruction---which accounts for the remaining gap between variant~\MuGauNormOnly\ ($\MuGauNormOnlyNUM$) and the blue curve ($\sim5.2$--$5.6\times10^{-6}$). Both effects (including subleading shape systematics and using the full \texttt{pyhf} framework) act in the same direction: they lower the $\Ba$ limit, making the bound of \refcite{\us} conservative. The factor-of-four improvement of \refcite{\us} over \refcite{\gaertner} is therefore a robust lower bound on the true sensitivity gain from working in the fine-grained $\qqrec$ space.

\medskip
{\bf Test 5: $\mcB_{\nunu}$ likelihood profile within $\usGaertnerlike$ analysis~--~}%
A further instructive check is to set $\mcB_a = 0$ in the $\usGaertnerlike$ analysis and examine the profile likelihood as a function of $\mu_{\nunu}$, in analogy with the counterpart test of \refsec{sec:robustness}. In particular, we set the background-normalization error to either 1\% or 50\%, to allow for a direct comparison with that test. The result is shown in \reffig{fig:BKa_Gaertnerlike} (right panel), alongside the Belle~II experimental profile for comparison. Two features stand out. 

First, the $\usGaertnerlike$ profile is more optimistic than the experimental one---steeper and with a higher central value---because the shape systematics absent from the $\usGaertnerlike$ analysis are specifically important for the $\BKvv$ component: their omission reduces the uncertainty on $\mu_{\nunu}$ and artificially sharpens the profile.

Second, in contrast to what was observed in \refsec{sec:robustness}, the $\usGaertnerlike$ profile cannot reproduce the Belle~II experimental profile at any background-normalization error value: it remains sharper than the experimental one even at 50\%, and the three curves for 0\%, 1\%, and 50\% are themselves distinguishable (with 0\% the sharpest). This behavior reflects a key difference from the $\qqrec$-based approach. In $\qqrec$ space, the $\BKvv$ signal and background shapes are nearly degenerate, so the background-normalization prior is the primary handle on their separation, and 1\% closely reproduces the experimental profile. In the $\usGaertnerlike$ space, the $\BDTt$ axis provides genuine shape discrimination between $\BKvv$ and its backgrounds; as a result, the width of the Belle~II profile is primarily set by the 231 shape nuisance parameters, and no background-normalization value alone can reproduce it. The price of this shape separation is the loss 
of $\qqrec$ resolution that is the key figure of merit for $\BKa$.

\section{Conclusions} \label{sec:conclusions}

\noi The choice of kinematic variable space in recasting $\BKvv$ data as a search for narrow invisible signals has a decisive impact on sensitivity. This note has traced the origins of a factor-of-four difference in the $\mcB_a$ limits between two recent independent analyses of the same Belle~II data.

The central finding is that $\qqrec$ resolution is the key figure of merit. The published model-agnostic likelihood of \refcite{\abuagnostic} encodes only three $\qqrec$ bins---sufficient for the broad $\BKvv$ spectrum it was designed for, but inadequate for a narrow signal whose structured $\qqrec$ distribution is subsumed into a single bin dominated by $\BKvv$ background. Its second axis, $\eta(\BDTt)$, was trained to separate $\BKvv$ from background and carries little discriminating power for $\BKa$, whose sole distinguishing kinematic variable is $q^2$ itself. Both limitations reflect a more fundamental property of the reweighting method of \refcite{\abuagnostic}: that method requires the alternative theory to remain close to the smooth null distribution. For a narrow or zero-width invisible state, $\sigma_1(q^2)\propto\delta(q^2-m_a^2)$, which is maximally far from the smooth null---precisely the least favorable regime.

These analytical expectations are confirmed through five numerical tests. Tests~1 and~2 establish the robustness of the \refcite{\us} approach: the $\mcB_a$ limit is insensitive to the background-normalization uncertainty, confirming strong shape separation between the axion signal and the $\BKvv$ background in $\qqrec$ space; the $\BKvv$ profile likelihood, by contrast, depends sensitively on this error---0\% gives a profile that is too sharp, 50\% one that is too broad, and 1\% reproduces the experimental profile closely---independently validating the error estimate; the ``dual-probe'' feature---the mutual independence of the $\mcB_a$ bound and the size of new physics in $\mcB_{\nunu}$---is preserved throughout.

Test~3, a direct rerun of the \refcite{\gaertner} likelihood with its full set of 231 nuisance parameters, shows that shape systematics lower $\mcB_a$ by $\sim 20\%$, confirming that omitting them from \refcite{\us} makes its bound {\em conservative}. It further shows that freeing $\muvv$ entirely improves the limit by roughly a factor of two, confirming that the $\qqrec\times\eta(\BDTt)$ space is sensitive to the SM-likeness of $\BKvv$---thus losing the dual-probe property of the approach in \refcite{\us}.

Tests~4 and~5 use a dedicated $\usGaertnerlike$ analysis, which applies the \refcite{\us} framework to the $\qqrec\times\eta(\BDTt)$ space of \refcite{\gaertner}, while keeping the statistical method, signal model, and background treatment identical to those of \refcite{\us}. This cleanly isolates the kinematic-axes effect: the $\sim$factor-of-five degradation in $\mcB_a$ relative to \refcite{\us} is attributable solely to the change of kinematic axes. The direct rerun of Test~3 and the $\usGaertnerlike$ analysis of Tests~4 and~5 are therefore complementary: the former probes the full \refcite{\gaertner} likelihood; the latter isolates the kinematic-axes contribution.

The results of this comparison bring, we believe, insights that extend beyond the specific case studied here. Likelihoods defined in spaces dominated by BDT variables are of limited use for recasting measurements into signals whose shape differs appreciably from the signal the BDT was trained on. We therefore advocate that experimental collaborations publish likelihood projections in the space of physical variables alongside BDT-based likelihoods, so as to maximise the reinterpretability of their measurements for the broadest possible range of new-physics scenarios.

\acknowledgments

\noi This work has received funding from the French ANR, under contract ANR-23-CE31-0018 (`InvISYble'), that we gratefully acknowledge.

\bibliography{bibliography}

%
%
%
%

\end{document}